\documentclass[aps,preprint,english,prb]{revtex4}
\usepackage{graphicx}
\usepackage[T1]{fontenc}
\usepackage[latin1]{inputenc}
\usepackage{float}
\usepackage{graphicx}
\usepackage{verbatim} 
\usepackage{babel}
\usepackage{amsmath}

\begin{document}

\title{
Absolute free energies estimated by combining pre-calculated molecular fragment libraries
}
\author{Xin  Zhang$^*$}
\affiliation{Department of Physics $ \& $ Astronomy, University of Pittsburgh, Pittsburgh, Pennsylvania 15260}
\author{Artem B. Mamonov\footnote {These authors contributed equally.} 
  and Daniel M. Zuckerman\footnote{Electronic mail: ddmmzz+@pitt.edu} }
\affiliation{Department of Computational Biology, School of Medicine, University of Pittsburgh, Pittsburgh,  Pennsylvania 15213}

\date {\today}

\begin{abstract}
The absolute free energy --- or partition function, equivalently --- of a molecule can be estimated computationally using a suitable reference system.  
Here, we demonstrate a practical method for staging such calculations by growing a molecule based on a series of fragments.  
Significant computer time is saved by pre-calculating fragment configurations and interactions for re-use in a variety of molecules.  
We employ such fragment libraries and interaction tables for amino acids and capping groups to estimate free energies for small peptides.  
Equilibrium ensembles for the molecules are generated at no additional computational cost, and are used to check our results by comparison to standard dynamics simulation.  
We explain how our work can be extended to estimate relative binding affinities.
\end{abstract}

\maketitle

%%% common symbols defined here
\newcommand{\arr}{\rightarrow}
\newcommand{\cref}{C^{\circ}}
\newcommand{\df}{\Delta F}
\newcommand{\dfkp}{\df_{k \arr \mathrm{phys}}}
\newcommand{\dfrp}{\df_{\mathrm{ref \arr phys}}}
\newcommand{\dfaa}{\df_{\mathrm{Ala \arr Ala}}}
\newcommand{\dfaat}[2]{\df_{\mathrm{Ala{#1} \arr Ala{#2}}}}
\newcommand{\dfan}{\df_{\mathrm{Ala \arr Nme}}}
\newcommand{\dfafn}{\df_{\mathrm{Ala4 \arr Nme}}}
\newcommand{\dfca}{\df_{\mathrm{Ace \arr Ala}}}
\newcommand{\dfcao}{\df_{\mathrm{Ace \arr Ala1}}}
\newcommand{\dfcan}{\df_{\mathrm{Ace-Ala \arr Nme}}}
\newcommand{\dfnon}{\df_{\mathrm{nonbonded}}}
\newcommand{\face}{F_{\mathrm{Ace}}}
\newcommand{\faceala}{F_{\mathrm{Ace-Ala}}}
\newcommand{\fala}{F_{\mathrm{Ala}}}
\newcommand{\fnme}{F_{\mathrm{Nme}}}
\newcommand{\fref}{F^{\mathrm{ref}}}
\newcommand{\fphys}{F^{\mathrm{phys}}}
\newcommand{\lav}{\left \langle}
\newcommand{\lb}{\left [}
\newcommand{\lcb}{\left \{}
\newcommand{\lp}{\left (}
\newcommand{\neff}{N_{\mathrm{eff}}}
\newcommand{\rav}{\right \rangle}
\newcommand{\rb}{\right ]}
\newcommand{\rcb}{\right \}}
\newcommand{\rp}{\right )}
\newcommand{\unit}{\mbox{1 \AA}}
\newcommand{\uphys}{U^{\mathrm{phys}}}
\newcommand{\uref}{U^{\mathrm{ref}}}
\newcommand{\ureflc}{u^{\mathrm{ref}}}
\newcommand{\xbf}{\mathbf{x}}
\newcommand{\zphys}{Z^{\mathrm{phys}}}
\newcommand{\zref}{Z^{\mathrm{ref}}}

\section{Introduction}
The use of a reference system for free energy calculations has a long history in physics and chemistry.~\cite{Frenkel-book}
The basic idea is to employ a reference system (``ref'') for which the absolute free energy is available, and which is as similar as possible to the physical system of interest (``phys'').  
Historically, Stoessel and Nowak applied the reference-system strategy to a molecular system for the first time, using a solid harmonic reference system in Cartesian coordinates.~\cite{Stoessel-1990}  Zuckerman and Ytreberg extended that work in two ways designed to improve overlap between the reference and physical systems:~\cite{Zuckerman-2006d} (i) by using internal coordinates; and (ii) by using a more flexible, numerically exact reference system based on histograms from a short dynamics simulation, rather than an artificial analytically tractable reference state.  Huang and Makarov also employed the reference-system approach embodied in (1), but in a different way.~\cite{Huang-2006}  
\begin{equation}
\fphys = \fref + \df_{\mathrm{ref \arr phys}} 
 \; ,
\label{f-from-ref}
\end{equation}
where $F^{\mathrm{x}}$ is the absolute free energy of model ``x,''  and $\dfrp$ is the free energy difference between the systems.  
In essence, this paper is about practical choices for the both the reference system and strategies for calculating  $\dfrp$ when the physical system is a large molecule.

Historically, Stoessel and Nowak applied the reference-system strategy to a molecular system for the first time, using a solid harmonic reference system in Cartesian coordinates.~\cite{Stoessel-1990}  Zuckerman and Ytreberg extended that work in two ways designed to improve overlap between the reference and physical systems:~\cite{Zuckerman-2006d} (i) by using internal coordinates; and (ii) by using a more flexible, numerically exact reference system based on histograms from a short dynamics simulation, rather than an artificial analytically tractable reference state.  Huang and Makarov also employed the reference-system approach embodied in (1), but in a different way.~\cite{Huang-2006}  

Other efforts to calculate absolute free energies for molecular systems have been ongoing for years in the groups of Meirovitch~\cite{Meirovitch-1992, Meirovitch-1999,  Cheluvaraja-2004, White-2004a} and Gilson~\cite{Gilson-1997,Chang-2003,Chang-2004,Chen-2004} and more approximately using harmonic and quasi-harmonic methods.~\cite{Go-1969} The work of Meirovitch builds on long-standing polymer-growth methodologies for estimating partition functions which date to the work of Rosenbluth and Rosenbluth.~\cite{Rosenbluth-1955} The original Rosenbluth work was generalized for higher efficiency and more realistic models by many workers.~\cite{Wall-1959,Meirovich-1982,Meirovich-1985,Garel-1990,Garel-1991,Velikson-1992,Bascle-1993,Grassberger-1993a,Grassberger-1995,Grassberger-1997,Bastolla-1998,Liu-2002,Liu-2007}  Ideas from these polymer-growth sampling methods also inform the present work.  

The present paper significantly extends the previous work by Ytreberg and Zuckerman~\cite{Zuckerman-2006d} by estimating absolute free energies for molecules built up gradually from molecular fragments.  Larger molecules can be treated, compared to our previous paper~\cite{Zuckerman-2006d}, because a series of staged intermediate systems are adopted.  In essence the free energy difference   of Eq. (1) is sub-divided (staged) into a sum of easy-to-calculate terms.  Staging increments are highly tunable, based on the choice of fragment sizes and even by selection of subsets of interactions as detailed below. The use of fragments in other types of molecular mechanics calculations has a long history.~\cite{Gibson-1987,Karplus-1991,Leach-book}

A key contribution of this work is a practical strategy of pre-calculation which minimizes the number of energy terms which need to be computed at each stage.  Specifically, for each fragment, a statistical library --- i.e., an ensemble  of configurations and their energies --- is stored; we have also used such libraries in Monte Carlo sampling~\cite{lbmc-2008}.  Additionally, for each covalently bonded fragment pair, we store the full interaction energy (based on all atoms) for every possible pair of configurations.  Such storage is quite practical on typical modern computers with > 1 GB of RAM.  During production simulations it is only necessary to compute interactions between fragments separated by one or more other fragments.  Needless to say, the stored libraries and interaction tables can be re-used in future simulations of the same or different molecules. The pre-calculation strategy, which has early conceptual roots~\cite{Wall-1957,Alexandrowicz-1969,Macedonia-1999}, appears to represent a significant practical advance over earlier polymer-growth calculations. The use of non-statistical libraries has been popularized in the Rosetta protein folding program.~\cite{Baker-2004}

There is substantial flexibility in the division of a molecule into fragments.  We have used single amino acids as fragments in this study, but larger segments and even different interaction subsets  as detailed below -- may also be practical.  The fragment-based approach could also be used to study protein-ligand binding, by growing small molecules into receptor binding pockets and estimating the free energies.  This can be seen as a statistical mechanical generalization of fragment-based ideas developed earlier.~\cite{Leach-book,Karplus-1991}

Our results, which employ single amino-acid fragments, are extremely encouraging.  The data indicate that absolute free energies for small peptides can be calculated rapidly and reliably.  Specifically, high-precision free energy estimates, with fluctuations of $\sim $ 0.3 kcal/mole, are obtained for 52-atom tetra-alanine in less than an hour of single-processor computing time, with a simple dielectric "solvent". We check our data by comparing the equilibrium ensembles (obtained simultaneously with the free energy estimates) with independent Langevin simulations.  As a further check, in one case, the free energy results are verified by an independent calculation using different fragments.
	
The remaining sections of the paper describe the methods, results, and our conclusions.  Our methods section provides full details for performing the calculations, including the generation of fragment libraries and interaction tables.  We also correct a technical error in our earlier study.~\cite{Zuckerman-2006d}  Our results describe both the free energy values and the analysis of the equilibrium ensembles.  Our discussion section describes possible improvements to the method and extension to the estimation of relative binding affinities using absolute free energies.
\section{Methods}

Our basic approach is to calculate the free energy of the physical system of interest based on the difference from a known reference system, as in Eq.\ (\ref{f-from-ref}), and also to stage the calculation using molecular fragments.The fragments not only permit the gradual staging of the calculation but also a tremendous savings of computer time based on the storage of (i) fragment configurations, (ii) energies internal to each fragment configuration, and (iii) interaction energies between covalently bonded fragments.
The low cost and high precision of the resulting estimates suggests we are far from the practical limit of the approach in the present implementation. However, a number of improvements to the implementation appear to be within easy reach, as described in our Discussion. 
All fragment libraries used in the present calculations are available at our website (www.ccbb.pitt.edu/Zuckerman).
\subsection{Model and systems}
All calculations employ a standard atomistic forcefield, OPLS-AA~\cite{Jorgensen-1996} at $T=298K$.
In the present report, our fragments are individual amino acids and capping groups. For simplicity in this initial investigation, we model the solvent by a simple uniform dielectric constant $\epsilon = 60$. We compute free energy estimates for alanine dipeptide (Ace-Ala-Nme), di-alanine (Ace-(Ala)$_2$-Nme), and tetra-alanine (Ace-(Ala)$_4$-Nme). Following standard conventions, Ace is Acetyl (CH$_3$-CO), Ala is Alanine (NH-CH(CH$_3$)-CO), and Nme is N-methylamide (NH-CH$_3$).

\subsection{A simple example}
\label{sec:ala-di}
Consider the calculation of the configurational free energy of alanine dipeptide based on a division into three fragments (Ace, Ala, Nme) which can be denoted (A, B, C) respectively (see Fig.1).
\begin{figure}[h]
\includegraphics[totalheight=3.5in]{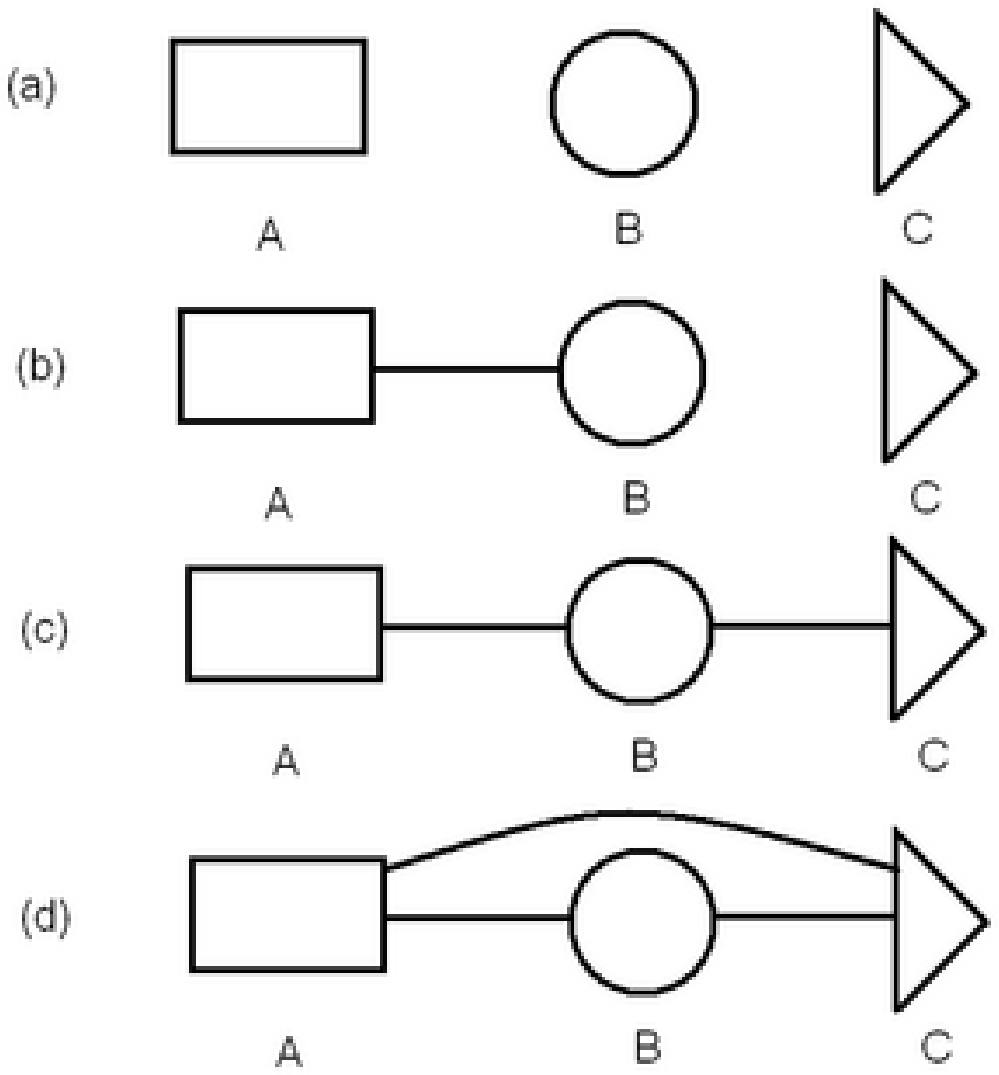}
\caption{
\label{fig:abc}
}
\end{figure} 
In advance, we calculate statistical libraries of configurations for each fragment, which are constant-temperature OPLS-AA ensembles based only on the atoms within the given fragment.
The libraries additionally include the six degrees of freedom necessary for joining the fragments, based on the use of ``dummy atoms'' as described below.
During the library generation process, the absolute free energy for each fragment is also calculated using a reference system as described previously.~\cite{Zuckerman-2006d}
A typical library will contain 10,000 configurations.
We also pre-calculate every possible interaction energy between covalently bound fragments --- i.e., a table of $10^8$ interaction energies for the A-B and B-C fragment pairs.

The calculation proceeds as schematized in Fig.\ \ref{fig:abc}, where the presence of a line connecting two fragments indicates that all interactions between the fragments is included.
The reference system (not shown) consists of fully independent coordinates, so that the fragments are not yet constructed. 
The first intermediate consists of the three non-interacting fragments, which include, however, all interactions \emph{within} each fragment.
Thus the fragment free energies, which are calculated and stored in advance, properly include the interactions among all degrees of freedom \emph{internal} to each fragment.
Other interactions are added in three stages: A-B interactions first, followed by B-C, and completed by A-C couplings.

In the first intermediate stage, the absolute free energies for the individual fragments are retrieved from disk.
(They are initially calculated following reference~\cite{Zuckerman-2006d} as detailed below.)
Next, A-B interactions are added by a standard free energy difference calculation.
Specifically, an ensemble of non-interacting A-B configurations is generated by random combination of fragments from the A and B libraries, and the resulting energy change is exponentially averaged in the usual way --- via Eq. (6) below.
The energy changes due to the combination do not need to be calculated in our scheme, however, because they have been tabulated in advance.
Additionally, the now interacting A-B fragments are ``resampled''~\cite{Liu-book} 
to correspond to the full potential for all degrees of freedom in \emph{both} fragments.
The details of resampling are given below --- see Eq.\ (\ref{resample}) --- but the bottom line is that one obtains 10,000-configuration ensemble of the partially grown molecule consisting of the A-B fragments.

The calculations then proceeds as if there are two fragments, A-B and C.
The two libraries are joined combinatorially \emph{but only accounting for the B-C interactions at this stage}.
The A-C interactions will be handled at a later stage.
Once again, the free energy change is calculated and the ensemble is resampled to reflect B-C interactions.
The resulting ensemble contains 10,000 configurations of the full molecule reflecting all interactions except those between fragments A and C.

In the final stage sketched in Fig.\ \ref{fig:abc}, the A-C interactions are added in a standard free energy difference calculation based on the the ensemble of the previous stage.
However, a standard energy call (for the whole molecule) is not required to save CPU time. Rather, our code only computes energy terms specific to the A and C fragments --- i.e., electrostatic and van der Waals interactions between atoms of A and those of C.
Once the energy changes have been obtained, the full free energy is rapidly estimated.
Resampling into the fully interacting ensemble can also be performed rapidly without additional energy calculations.

It is not difficult to imagine generalizing this example to systems with more fragments.
It is also worth noting that, strictly speaking, the final stage was not necessary.
That is, we could have added the A-C interactions simultaneously with the B-C combination since the full molecular configurations were constructed at that point.
These choices illustrate the flexibility intrinsic to staging the calculation with fragments, as we detail further in our Discussion.
Additional staging flexibility results, of course, from the initial choice of the fragments --- i.e., smaller fragments lead to staging in finer increments.

\subsection{Basic formalism}
Our calculation of the absolute free energy $\fphys$ for a molecule divided into fragments is based on standard, straightforward equations.
The only novel aspect of the formalism is our particular choice of stages based on the addition of fragments and/or inter-fragment interactions.
Although our heavy reliance on pre-calculated information has very significant practical implications, it does not affect the formalism.

\subsubsection*{Definition of the free energy}
The fundamental object of interest is the absolute classical free energy $\fphys$ for an implicitly solvated molecule.
The molecule is taken to consist of $N$ atoms, and its \emph{internal}-coordinate configurations are denoted by $\xbf$.
The potential energy function will be a standard forcefield (here, OPLS-AA~\cite{Jorgensen-1996}), possibly augmented by an implicit solvation model;
the full potential energy including any solvation will be denoted by $\uphys(\xbf)$.
The free energy, which is a functional of $\uphys$, is defined by the dimensionless \emph{configurational} partition function at temperature $T = 1/k_B \beta$ via
\begin{equation}
\fphys[\uphys] = -k_B T \ln \lcb
  \frac{1}{\lp \unit \rp^{3N-3}} \int d\xbf \,
   e^{-\beta \uphys(\xbf)}
\rcb \; ,
\label{fphys-units}
\end{equation}
where the measure of integration $d\xbf$ is understood to include any necessary Jacobians. Kinetic energy terms have already been integrated out.
Both the dimensionless character of the partition function in Eq.\ (\ref{fphys-units}) and the angstrom-based normalization result from a particular choice for the standard concentration $\cref$ (defined in references~\cite{Chang-2003,Zuckerman-2006d}) --- or equivalently, for the volume containing our implicitly solvated molecule.
In particular, we have chosen
$\cref \equiv 8\pi^2 ( \unit )^{3N-3} Q_p / \sigma$,
where $Q_p = \prod_{i=1}^N  ( 2 \pi \, m_i \, k_B T / h^2 )^{3/2}$ results from the momentum integrals, $h$ is Planck's constant, and $\sigma$ is the molecule's symmetry number.
We note that our chosen standard concentration varies based on the molecule (i.e., based on the number and masses of its atoms), and also eliminates the temperature dependence of $Q_p$.
However, in almost every application of interest (see Discussion, below), the absolute free energy calculated here ultimately will be used to estimate a free energy difference and eliminate any artifacts due to $\cref$.

Our single-molecule formulation, as noted, allows for ``implicit solvation'' using an effective solvent term in $\uphys$ that is solely a function of the internal molecular coordinates $\xbf$.
The present calculations employ a simple uniform dielectric constant ($\epsilon = 60$).
In our Discussion, we address the minor technical issues involved with using a more realistic implicit solvent model. 

One issue of dimensionality is worth emphasizing.
Although there are $3N-6$ internal coordinates for a molecule consisting of $N$ atoms, the integral of Eq.\ (\ref{fphys-units}) has dimensionality of length to the power $3N-3$.
This is because $N-1$ bond lengths remain in the full set of internal coordinates $\xbf$, each of which contributes three powers of length.
Put another way, of the six excluded rigid-body/center-of-mass coordinates, the three orientation angles are dimensionless; more specifically, the angles integrate to the factor of $8\pi^2$ included in the definition of $\cref$

\subsubsection*{Staging the free energy calculation}
As illustrated in the example of Sec.\ \ref{sec:ala-di}, we will calculate the free energy in a series of stages.
These can be understood most easily by adding and subtracting the free energies corresponding to $k$ intermediate models, 
\begin{eqnarray}
\fphys & = & \lp \fphys - F_k \rp + \lp F_k - F_{k-1} \rp + \cdots + 
         \lp F_1 - \fref \rp + \fref
\\
& = & \dfkp + \df_k + \cdots + \df_1 + \fref \; ,
\label{f-staged}
\end{eqnarray}
where $\df_j = F_j - F_{j-1}$  and $F_j[U_j]$ is defined in analogy to Eq.\ (\ref{fphys-units}) for the intermediate models defined by $U_j$. 
The $U_j$ potentials will be specified below.

All free energy difference calculations will be performed here using the ``perturbation'' formulation.~\cite{Zwanzig-1954}
Explicitly, for two arbitrary potential energy functions $U_a$ and $U_b$, one has
\begin{eqnarray}
\df_{a \arr b} = F_b[U_b] - F_a[U_a] 
& = & -k_B T \ln \lav \exp \lb -\beta \lp U_b - U_a \rp \rb
                 \rav_a
\\
& \simeq & -k_B T \ln \lcb \frac{1}{N_a}
             \sum_{i=1}^{N_a} \exp \lb -\beta \lp U_b(\xbf_i) - U_a(\xbf_i) \rp \rb
             \rcb
\label{df-est}
\end{eqnarray}
where the subscript $a$ denotes an average performed over configurations distributed according to the $U_a$ ensemble and $N_a$ is the number of configurations in that ensemble.
Eq.\ (\ref{df-est}) is used to estimate the free energy differences required in Eq.\ (\ref{f-staged}), and it is exact in the limit $ N_a \rightarrow  \infty $.

We emphasize that succeeding intermediates are constructed to have progressively narrower distributions as more interactions are added, as in the alanine dipeptide example.
In other words, we ensure good overlap and reliable $\df$ estimates by proceeding in the generalized ``insertion'' direction.~\cite{Peter1997,Kofke1998}

\subsubsection*{Resampling to obtain staged equilibrium ensembles}
As our calculation proceeds through the various stages, we will require the correspoinding equilibrium ensembles for each stage, primarily for use in Eq.\ (\ref{df-est}).
These are obtained by ``resampling,'' the process of converting an ensemble for one distribution into another by eliminating, duplicating, and/or adjusting the weights of configurations in the original distribution.~\cite{Liu-book}
In our case, we primarily use elimination of configurations from a larger ensemble (e.g., all combinations of fragments A and B) to create a smaller one (e.g., the interacting A-B ensemble); we do not adjust weights.
More specificially, to resample an ensemble of configurations $\xbf_a$ generated according to $U_a$ into a $U_b$ ensemble, the original configurations are randomly selected with probability proportional to the ratio of Boltzmann factors, 
\begin{equation}
\label{resample}
e^{-\beta[U_b(\xbf_a) - U_a(\xbf_a)]} \; .
\end{equation}
Operationally, we select configurations by forming a cumulative distribution function (cdf) based on the normalized set of ratios (\ref{resample}), and then choosing from this cdf as many times as desired.

\begin{figure}[h]
\includegraphics[totalheight=4in]{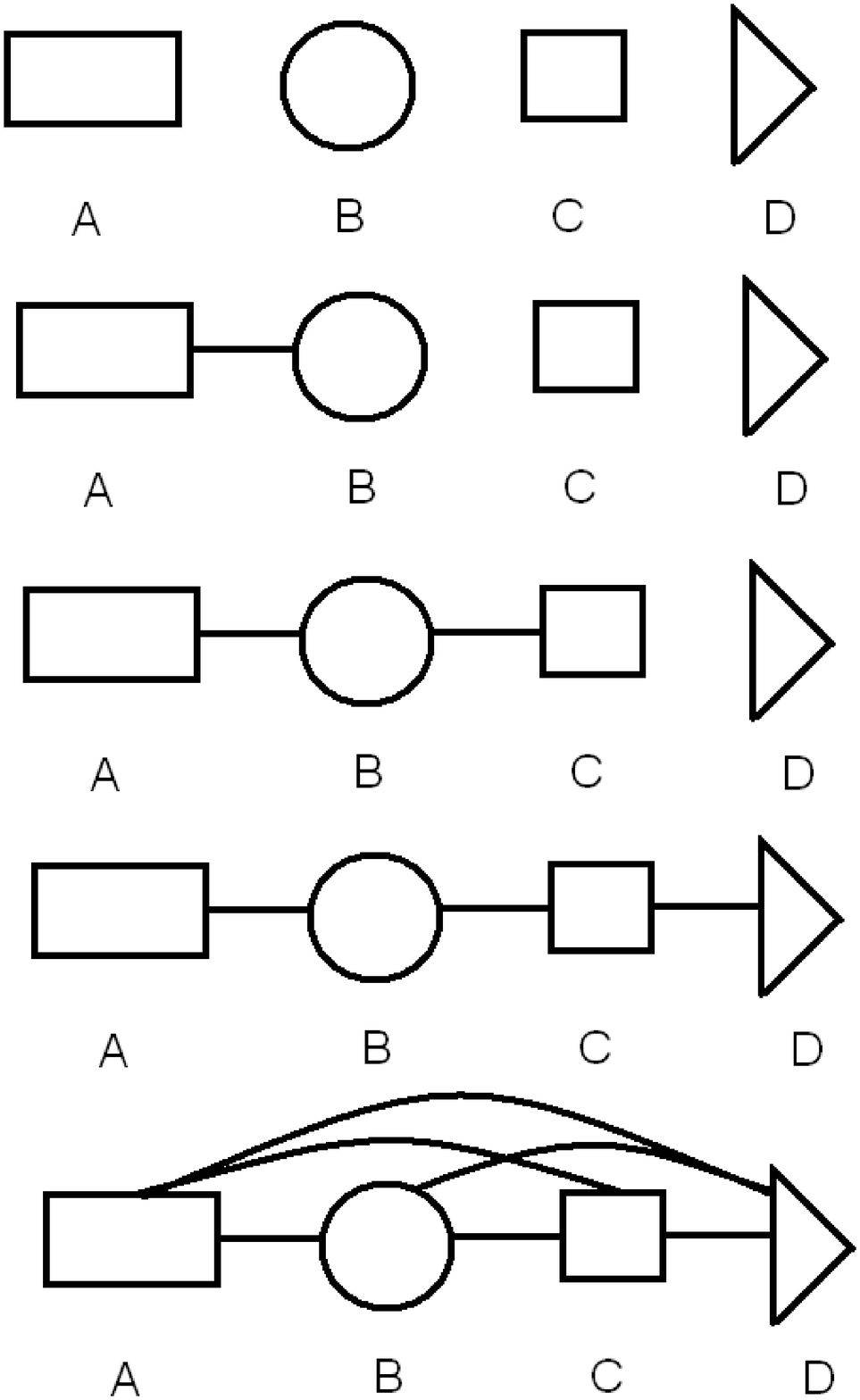}
\caption{
\label{fig:abcd}
}
\end{figure} 

\subsection{Choice of intermediate models}
\label{sec:staging}

As already noted, the set of intermediate models $\{U_j\}$ can be chosen in a variety of ways.
In the present study, we employ $k$ intermediates for a molecule divided into $k$ fragments.
This was exemplified for alanine dipeptide, which is divided into three fragments.

The present study uses a uniform staging strategy for all molecules examined, as exemplified in Figs.\ \ref{fig:abc} and \ref{fig:abcd}.
The reference system consists of \emph{all} coordinates fully independent --- both within and between fragments --- as in our previous work.~\cite{Zuckerman-2006d}
The first intermediate stage adds interactions \emph{within fragments}, so that one has true molecular fragments but no interactions between fragments.
We then add interactions between neighboring, covalently bound fragments --- i.e., among all the atoms in the neighboring fragment pair --- one fragment pair at a time.
The final stage of this simple scheme involves the addition of all remaining interactions, which occur solely between non-adjacent fragments.
The result is a molecule with atoms interacting fully according to a standard forcefield and possibly continuum solvent model. 

Because interactions among previously non-interacting components are added at every stage, it is expected that the configuration space will become increasingly narrow.
Such a progressive narrowing justifies the use of the perturbation expression (\ref{df-est}).

To explicitly illustrate the staging scheme employed here, consider the case of a molecule divided into the three fragments A, B, and C, as in Fig.\ \ref{fig:abc}.
We denote by $\ureflc_i$ the reference potential for internal coordinate $x_i$, where the full set is $\xbf = (x_1, x_2, \ldots )$.
For the fragments, we let $U_y$ be the potential energy for all interactions internal to fragment $y$, and $U_{yz}$ is the potential energy for all interactions between the $y$ and $z$ fragments.
A three-fragment molecule would be staged as follows:
\begin{equation} 
\begin{gathered}
\uref = \sum_{i=1}^{3N-6} \ureflc_i(x_i)\\
U_1 = U_A+U_B+U_C   \\
U_2 = U_1 + U_{AB} \\ 
U_3 = U_2 + U_{BC}  \\
\uphys = U_3 + U_{AC} \; .
\end{gathered}
\label{three-stages}
\end{equation}

The choice of the reference potentials $\{\ureflc_i\}$ is guided by the forcefield, as detailed below in Sec.\ \ref{sec:libraries}.

A four-fragment molecule, such as di-alanine (Ace-(Ala)$_2$-Nme) schematized in Fig.\ \ref{fig:abcd}, would be staged according to:
\begin{equation} 
\begin{gathered}
\uref = \sum_{i=1}^{3N-6} \ureflc(x_i)\\
U_1 = U_A + U_B + U_C + U_D  \\
U_2 = U_1 + U_{AB} \\ 
U_3 = U_2 + U_{BC}  \\
U_4 = U_3 + U_{CD} \\
\uphys = U_4 + U_{AC} + U_{BD} + U_{AD} \; 
\end{gathered}
\label{four-stages}
\end{equation}
As described in the Discussion, it is also possible to stage the final (``non-bonded'') pairwise interactions separately. 

We anticipate that significant optimization can be obtained by adjusting fragmentation and staging schemes.
While our Discussion, below, describes more gradual staging strategies, the present initial report is limited to the single staging strategy given above.

\subsection{The non-interacting reference system}
The computation of the (absolute) reference free energy $\fref$ is perhaps the most technically involved step of the calculation.
The remaining free energy differences in the decomposition of $\fphys$ in Eq.\ (\ref{f-staged}) are estimated using a simple, standard method.
For the reference free energy, however, great care must be taken with the normalization and Jacobian factors of the chosen probability distributions.
Indeed, our previous report~\cite{Zuckerman-2006d} includes an error in this regard, as explained at the end of this subsection.

As described in our discussion of staging (Sec.\ \ref{sec:staging}), the reference system for all molecules studied here consists of the set of non-interacting internal coordinates.
The reference potential energy function will be constructed, following our previous work~\cite{Zuckerman-2006d}, so that the reference partition function is normalized to one.
That is, we will \emph{construct} our reference model $\uref$ so that 
\begin{equation}
\zref[\uref] = 
e^{-\beta \fref} = 
  \frac{1}{\lp \unit \rp^{3N-3}} \int d\xbf \,
   e^{-\beta \uref(\xbf)}
   \equiv 1 \; ,
\label{zref-one}
\end{equation}
where the same standard concentration as in Eq.\ (\ref{fphys-units}) has been used implicitly.
(From this point forward, we will omit writing the length units, but they should be implicitly associated with every bond-length integration.)
The motivation for the unit normalization of $\zref$ is that application of a logarithm leads to the simplifying value,
\begin{equation}
\fref \equiv 0 \; ,
\label{fref-zero}
\end{equation}
for \emph{every} system.

While there are many ways to construct $\uref$ to satisfy the required normalization of Eq.\ (\ref{zref-one}), we use the strategy of employing independent internal coordinates as in our earlier work.~\cite{Zuckerman-2006d}
As usual, the full set of $3N-6$ internal coordinates,
$\xbf = (x_1, x_2, \ldots, x_{3N-6})$ consists of $N-1$ bond lengths, $N-2$ bond angles, and $N-3$ dihedrals.
So long as the distribution of each individual coordinate is normalized when integrated with the appropriate Jacobian factor $J$, the full distribution will be normalized.
%Equivalently, as will be done in this report, one can use the ``Jacobian-embedded'' coordinates, 
%$x_i = (r_i^3/3)$, $x_i = -\cos \theta_i$, or $x_i = \omega_i$, for a bond length, angle, or dihedral, respectively.~\cite{Zuckerman-2006d}

Because total reference energy is given by a simple sum of independent terms,
\begin{equation}
\uref(\xbf) = \sum_{i=1}^{3N-6} \ureflc_i(x_i) 
\label{uref}
\end{equation}
the desired normalization (\ref{zref-one}) is ensured by enforcing
\begin{equation}
\int dx_i \, J(x_i) e^{-\beta \ureflc_i(x_i)} = 1 \; ,
\label{norm-coord}
\end{equation}
where the inclusion of inverse length units is understood for bond-angle integrals.
In words, then, each individual potential $\ureflc_i$ must include suitable normalization --- which is accomplished by offsetting the potential by the log of the integrated (un-normalized) Boltzmann factor. See reference~\cite{Zuckerman-2006d} for further information.

(As detailed in Sec.\ \ref{sec:libraries}, peptide $\phi$ and $\psi$ angles were, in fact, sampled together from a single distribution based on a pairwise energy function $\ureflc_{\phi\psi}$.
These angles are independent from all other coordinates, however.
We emphasize that this exception does not alter the basic formalism, which has been simplified very slightly for clarity.)

It is very useful to observe that normalization of the coordinate distributions via Eq.\ (\ref{norm-coord}) can be achieved either using standard analytic forms --- e.g., Gaussians --- or via numerical histogramming procedures.~\cite{Zuckerman-2006d}
Thus, there is great flexibility in the choice reference distributions embodied in the reference potentials.
In addition to forcefield terms, prior knowledge, such as from a simulation, can be used in constructing the set $\{ \ureflc_i \}$. The reference potentials chosen for the present study are described below in Sec.\ \ref{sec:libraries} on library construction. 

One word of warning is appropriate here.
Although it is possible to describe the internal configuration of a molecule using additional bond angles to substitute for dihedrals in some cases, the Jacobian for such a description appears \emph{not} to be well-defined.
Therefore, it is necessary to use the standard description with $N-2$ bond angles and $N-3$ dihedrals.
Unfortunately, we were unaware of this point during our original study,~\cite{Zuckerman-2006d} and therefore an erratum will be prepared correcting the resulting numerical errors.

\subsection{First intermediate: non-interacting fragments}

The first intermediate stage adds only localized interactions to the non-interacting reference model, as illustrated in Figs.\ \ref{fig:abc} and \ref{fig:abcd}.
Specifically, once a molecule is divided into fragments (A, B, C, ...), the first intermediate includes only interactions occuring \emph{within fragments}. 
The fragments exactly divide all coordinates so that we can write
$\xbf = (\xbf_A, \xbf_B, \ldots)$
and the potential energy function for this stage is given by
\begin{equation}
U_1(\xbf) = U_A(\xbf_A) + U_B(\xbf_B) + \cdots \; ,
\label{u-one}
\end{equation}
where $U_y$ includes all interactions from the \emph{full forcefield} (OPLS-AA, in our case) among the fragment coordinates $\xbf_y$ for $y = A, B, \ldots$. 
Importantly, the fragment potential $U_y$ includes all non-bonded interactions --- electrostatic, van der Waals --- among the atoms of the fragment.
(Sec.\ \ref{sec:libraries} on our libraries describes the treatment of connecting ``dummy'' atoms.)

The free energy for this stage --- i.e., $F_1[U_1]$ for use in the key equation (\ref{f-staged}), recalling $\fref \equiv 0$ --- can be calculated by using the standard perturbation relation (\ref{df-est}).
For such a computation, one would use $U_a = \uref$ and $U_b = U_1$ along with an ensemble distributed according to the Boltzmann factor of $\uref$.

In practice, \emph{once the libraries are generated, no calculation of energies needs to be done.} 
As detailed below the libraries are generated (just once, for repeated use in many systems) based on the $\uref$ distribution.
Thus, during the library generation process, it is a trivial matter to calculate the absolute free energy for each fragment using Eq.\ (\ref{df-est}).
Thus, individual fragment free energies $F_y$ are calculated in advance that exactly sum to the desired first-stage free energy: 
\begin{equation}
F_1 = F_A + F_B + \cdots \; .
\label{f-one}
\end{equation}
Further, the independent-coordinate distributions are subsequently resampled based on Eq.\ (\ref{resample}) to generate the library distributions --- i.e., ensembles for the $U_y$ Boltzmann factors --- for use in subsequent stages.

\subsection{Construction of fragment libraries}
\label{sec:libraries}

As just described, fragment libraries are critical to the calculation of the free energy of the first intermediate stage, $F_1$.
The libraries also greatly facilitate computations for the rest of the intermediates.

In general terms, fragment configurations are generated by sampling internal coordinates according to the independent probability distributions which constitute the reference system. 
The generated configurations are then used to calculate fragment free energies, $F_y$ for $y = A, B, \ldots$.
The configurations are also reweighted into an ensemble distributed according to the full forcefield for $\xbf_y$, the degrees of freedom internal to fragment $y$. 
Typically, such a procedure can be applied only to systems with a sufficiently small number of degrees of freedom. 
For large systems with enough correlated degrees of freedom, there tends to be insufficient overlap with the reference system of independent coordinates.
That is, only a tiny fraction of the reference-distributed configurations will be important in the interacting ensemble.
 Therefore, the choice of the generating probability is essential for the efficient generation of libraries.

We found that for fragments the size of alanine residues, rather simple probability functions were sufficient for generating tens of thousands of (statistically independent) configurations in weeks of single-CPU time. 
This is a negligible cost because once a library is generated, it can be used in multiple simulations. 

Different coordinate types are best sampled with different distributions, as is suggested by the forcefield terms.
Regardless of the particular choice, the specification of the distribution immediately implies the functional form for the reference potential $\ureflc_i$ from Eq.\ (\ref{norm-coord}). 
We found that simple Gaussian distributions, with parameters extracted from a short Langevin simulation, worked well for bond lengths and bond angles.
For ``stiff'' dihedrals, such as those in relatively planar groups (e.g., peptide bond), a Gaussian is also appropriate. 
For ``soft,'' rotatable dihedrals ---  such as $\phi$, $\psi$ and $\chi$ angles in amino acids --- we simply extracted histograms from a Langevin simulation of alanine dipeptide, as described in reference~\cite{Zuckerman-2006d}.
A two-dimensional (correlated) probability function was used for the ($\phi,\psi$) dihedral pair, but a one-dimensional distribution was used for all other dihedrals.
 
Based on the distributions just described, internal coordinates were sampled independently (except for pairwise sampling of $\phi$ and $\psi$ dihedrals) using an in-house program written in C. 
Generated configurations were saved to disk and converted to Cartesian coordinates. 
The corresponding forcefield energies for each configuration were calculated using the ``analyze'' module of the Tinker software package.~\cite{Ponder-1987} 
Based on these values and the known reference energies, the individual fragment free energies were calculated using Eq.\ (\ref{df-est}).
A simple resampling procedure~\cite{Liu-book} was used to generate a fragment ensemble distributed according to the forcefield; see Eq.\ (\ref{resample}).
Only a small fraction ($\gtrsim 10^{-4}$) of reference-ensemble configurations remain after resampling, requiring extensive sampling of the reference ensemble and weeks of CPU cost, as mentioned earlier. 

For this study, we generated libraries consisting of 10,000 configurations.
All fragment libraries were sampled according to OPLS-AA forcefield at $T=298K$, with a simple dielectric constant ($\epsilon = 60$) modeling solvent.
The choice of dielectric constant was motivated by the reasonable behavior observed in separate Langevin simulations of poly-alanine systems (data not shown). 

As noted earlier, for all possible $10^8$ \emph{covalent} (neighboring) pairings of fragments, we also tabulated the interaction energies from the forcefield, accounting for all atoms in the fragment pair.
Suitable corrections for dummy atoms (see below) were made.
In other words, for a simple two-fragment system, all interactions are stored.

\subsubsection*{Use of dummy atoms}
Because fragments are sampled independently from each other, the six degrees of freedom that specify the relative orientation of neighboring fragments are included with the fragments. 
For this purpose ``dummy'' atoms are used to provide the extra coordinates. 
We stress that our use of dummy atoms was implemented carefully to avoid adding additional degrees of freedom (e.g., certain bond lengths and angles). 
We chose to have the dummy atoms interact with the true fragment atoms for better overlap with subsequent ensembles. 
Thus, when the fragments are joined, the interaction energies of dummy atoms should be subtracted from the full fragment energy because dummy atoms are replaced with neighboring fragment atoms.
(Of course, it is simpler to have  non-interacting dummy atoms.)

The dummy atoms used at the N-terminus of a fragment are carbonyl C, carbonyl O and terminal alpha-C with valence set to one. 
The dummy atoms used at the C-terminus are amide N, amide H, and terminal alpha-C with valance also set to one. The dummy atoms were assigned the same forcefield parameters as used in the corresponding fragment atoms.
\subsection{The second and subsequent intermediates: adding neighboring fragment interactions}
\label{sec:neighbors}

Returning again to the scheme embodied in Eq.\ (\ref{f-staged}), as well as in Figs.\ \ref{fig:abc} and \ref{fig:abcd}, the next intermediates add interactions between neighboring fragments.
These can be considered the ``bonded'' interactions in the space of fragments, but non-bonded interactions among all atoms in the neighboring pair are included.
Explicitly, the models for the remaining intermediates are described by
\begin{eqnarray}
U_2(\xbf) & = & U_A(\xbf_A) + U_B(\xbf_B) + \cdots + \; U_{AB}(\xbf_A, \xbf_B) \\
%%%%%%%%%%%%%%%%%%%%%%%%%%%%%%%%%%%%%%
U_3(\xbf) & = & U_2(\xbf) + U_{BC}(\xbf_B, \xbf_C) + \cdots \; , 
\end{eqnarray}
where $U_{yz}$ is the full interaction energy --- based on the forcefield and solvent model --- between fragments $y$ and $z$.

Formally, it is clear what needs to be done.
The ensemble of the previous stage $j-1$ should be used to calculate $\df_j$ using the perturbation relation (\ref{df-est}) with $U_{j-1}$ and $U_j$.

Again, however, possession of the libraries and interaction tables leads to dramatic practical implications.
For instance, by construction, the energy $U_2 - U_1$ is simply the pre-stored energy $U_{AB}$; similarly $U_3 - U_2 = U_{BC}$.
These tabulated energies are used directly in Eq.\ (\ref{df-est}) without the need for additional energy calls.
The required ensembles for each stage are generated by the rapid resampling procedure of Eq.\ (\ref{resample}).
In this way, one readily generates the free energy differences $\df_2, \df_3, ... $ required for the evaluation of $\fphys$ via Eq.\ (\ref{f-staged}).

Caution is required when the molecule of interest contains repeated fragment pairs.
While the same libraries can be used for the repeats, say at intermediate stages $j$ and $m$, \emph{the corresponding values of $\df_j$ and $\df_m$ will be different in general}.
To see the reason, consider the case of the tetra-alanine peptide studied below.
The term $\df_2$ corresponds to including the interaction of the already combined fragments Ace-Ala with the next Ala.
Note that the free energy difference $\df_2$ is calculated via Eq.\ (\ref{df-est}) using the Ace-Ala ensemble as the ``a'' system.
By contrast, consider the calculation of $\df_3$ for the addition of the next Ala --- now to the Ace-(Ala)$_2$ ensemble.
Although the free energy change will be based upon the identical (tabulated) interactions, the associated Boltzmann factors in Eq.\ (\ref{df-est}) will be \emph{weighted differently --- i.e., occur with different frequencies --- due to the differing initial ``a'' ensembles.}
In turn, this will lead to different free energy changes, so that $\df_3 \neq \df_2$.

\subsection{The final free energy difference: non-neighboring interactions}
As described in the master scheme of Eq.\ (\ref{f-staged}) and illustrated in Figs.\ \ref{fig:abc} and \ref{fig:abcd}, the final calculation needed to obtain $\fphys$ entails the inclusion of all remaining interactions in the forcefield and solvent model.
These interactions, excluded until now, occur between atoms in non-neighboring pairs.
As described in Sec. II D, for a molecule of $k$ fragments, the full physical potential energy function (i.e., the forcefield) can be written as the difference from the final ($k$th) intermediate:
\begin{equation}
\uphys(\xbf) = U_k(\xbf) + \sum_{y..z} U_{yz}(\xbf_y, \xbf_z) \; ,
\end{equation}
where the sum is over non-neighboring pairs of fragments --- i.e., AC, AD, BD, ....

In this case, the necessary energy terms for use in the calculation of $\dfkp$ via Eq.\ (\ref{df-est}) must be calculated.
They cannot readily be stored in advance, due to the combinatorial explosion of possible configurations.
For instance, with libraries of $10^4$ configurations, there are $10^{12}$ possible configurations for three fragments, which is beyond the range of current commerical machines.

\subsection{Generating an equilibrium ensemble without additional energy calls}
\label{sec:equil}
The physical ensemble, distributed according to the Boltzmann factor of the forcefield, can be generated by resampling the $U_k$ ensemble --- the last intermediate --- using Eq.\ (\ref{resample}).
In this case, the ``a'' ensemble corresponds to $U_k$ and the ``b'' ensemble to the full forcefield and (implicit) solvent model.
Because all energy terms have already been calculated, no additional energy calls need to be made.
The necessary resampling computation is extremely fast compared with preceding stages of the protocol.

\subsection{Checking the code and estimating uncertainty}
\label{sec:method-check}
Although the formalism governing the present study is mostly straightforward, our in-house computer program not only needs to reproduce standard forcefield results, but also requires complicated ``dissections'' of various subsets of forcefield terms.
We therefore performed three types of checks on our code.
(i) We checked that the forcefield energy for full molecular configurations exactly reproduces the results reported in Tinker (data not shown).
This verifies that we have correctly accounted for our dummy-atom energy terms.
(ii) Using our previously developed ``structural histograms'' for analyzing configuration-space distributions,~\cite{Zuckerman-2006b,Lyman-2007} we checked that the equilibrium ensembles produced during our free energy calculations agree with independent Langevin simulations.
This data is shown in the Results section, and generated as explained below.
(iii) Finally, we performed a check to ensure that our final free energy values are independent of the choice of fragments.
This data, for two- and three-fragment decompositions of alanine dipeptides is also shown in the Results section.

\subsubsection*{Statistical error}
Statistical uncertainties were calculated by running 20 independent computations for every free energy value reported.
Twice the standard deviation among these 20 values is reported, which quantifies the scale of expected statistical error for a \emph{single} simulation.
The repeated simulations were run using 20 independent sets of libraries for the various fragments --- i.e., the calculation was started all the way at the beginning in each repeat.
However, because the overlap between various stages is the limiting factor in the quality of the free energy results, rather than our fairly large libraries, we anticipate similar error estimates would be obtained for one set of libraries.

\subsubsection*{Analyzing equilibrium ensembles/distributions}
%explain struct hist, ref structs, voronoi[ref]
In two previous studies,~\cite{Zuckerman-2006b, Lyman-2007} we have developed methods for comparing equilibrium distributions for molecular systems of arbitrary complexity.
The central idea is to employ a ``structural histogram'' which simply classifies (divides) configuration space into a number of ``bins'' (regions).
Two correct simulations should yield the same results for the fractional populations of the bins, within statistical error, regardless of whether the bins correspond to physical states/free energy basins. 
(Furthermore, the statistical uncertainty in the population estimates can be used to quantify the ``effective sample size''.)~\cite{Lyman-2007}
In the present work, we compare equilibrium distributions from fragment combination and from standard Langevin simulations based on structural histograms.
The particular histograms employed in the present study have five bins derived from a Voronoi construction;~\cite{Voronoi-1907}
the reference structures for the Voronoi procedure are derived from the equi-probability scheme described in reference.~\cite{Lyman-2007}
Although the resulting bins are not exactly equally probable, each is guaranteed to represent a contiguous region in configuration space due to the Voronoi construction.

\section{Results}
The absolute configurational free energy $\fphys$ was calculated for the monomer, dimer, and tetramer alanine peptides:
alanine dipeptide (Ace-Ala-Nme), di-alanine (Ace-(Ala)$_2$-Nme), and tetra-alanine (Ace-(Ala)$_4$-Nme).
For alanine dipeptide, the free energy was estimated based on two different fragment sets as a check on our code.
Twenty independent calculations for every $\fphys$ estimate were performed to quantify uncertainty, as described above.
Additionally, every free energy calculation also yields an equilibrium ensemble, which is compared to independent Langevin simulations.

The results are very positive in every regard, and rather rapid as reported at the end of the Results section.
The amount of memory used, which is a key to the present calculations, is also reported.

The results reflect the uniform protocol adopted here.
First, absolute free energies for non-interacting fragments are calculated.
Then free energy changes resulting from interactions among covalently bound fragments are added (``bonded'' terms, in the space of fragments), one at a time in sequence.
Finally, all remaining interactions are added, which account to (``non-bonded'') interactions among all non-sequential fragment atoms.
The final free energy values reflect the full OPLS-AA forcefield~\cite{Jorgensen-1996} as implemented in Tinker.~\cite{Ponder-1987}

\subsection{Alanine dipeptide using two different fragmentations}
Because of the complexity of the fragmentation procedure and the lack of reference standards for absolute free energy values, we wanted to ensure our code was introducing no artifacts.
We were particularly concerned about the interacting dummy atoms which introduce ``temporary'' energy terms, that must be corrected for properly at every combination stage.
We find excellent agreement between free energy estimates based on two- and three-fragment decompositions.

\subsubsection*{``Standard'' three-fragment decomposition}
In our standard decomposition for the present study, we separate peptide and amino acid groups.
For alanine dipeptide (AD), then, the three standard fragments are Ace, Ala, and Nme, and the corresponding stages for the free energy calculation are given in Eq.\ (\ref{three-stages}).
Recalling our convention that $\fref \equiv 0$, the free energy terms from Eq.\ (\ref{f-staged}) can be written as
\begin{equation}
\begin{gathered}
\fref \equiv 0 \\
\df_1 = \face + \fala + \fnme \\
\df_2 = \dfca \\
\df_3 = \dfan \\
\df_{3 \arr \mathrm{phys}} = \dfnon \; .
\end{gathered}
\label{ad-three}
\end{equation}
where $F_y$ is the absolute free energy (including dummy atoms) for fragment $y$ and $\df_{x \arr y}$ indicates the free energy change of combining fragments $x$ and $y$ (which includes all bonded and non-bonded terms, as well as the correction of dummy terms). 
Finally, $\dfnon$ denotes the free energy change in going from an ensemble where sequentially separated fragments do not interact to a fully interacting ensemble (in this case, the Ace-Nme interactions are added).

\subsubsection*{Two-fragment decomposition}
As an alternative decomposition, we used Ace-Ala as one fragment and Nme as the other.
Importantly, the Ace-Ala library and absolute free energy were \emph{not} generated from a combination of the two smaller libraries, but instead from a ground-up calculation based on independent coordinates as described in the Methods section.

In this case, then, the free energy terms from Eq.\ (\ref{f-staged}) become
\begin{equation}
\begin{gathered}
\fref \equiv 0 \\
\df_1 = \faceala + \fnme \\
\df_{1 \to phys} = \dfcan \; ,
\end{gathered}
\label{ad-two}
\end{equation}
where it is notable that in the two-fragment case, \emph{all} interactions are included in the libraries and interaction tables.
In other words, no energy calls at all are needed.

\subsubsection*{Comparison of free energies}
There is essentially perfect agreement between free energies estimate via the two independent decompositions, which provides a reassuring check on our computer program.
The full results are given in Table \ref{tab:ala-dipep}.
Notably, the two-fragment decomposition has a higher variance, which probably results from a decreased ``precision'' in the pre-generated Ace-Ala ensemble.
In the composite pre-generated Ace-Ala ensemble, the whole configuration space is represented by $10^4$ configurations, whereas when Ace and Ala from separate $10^4$-member libraries are combined, there is a much denser coverage of configuration space.

\subsubsection*{Equilibrium ensemble compared to standard simulation}
\begin{figure}
\includegraphics[totalheight=2.25in]{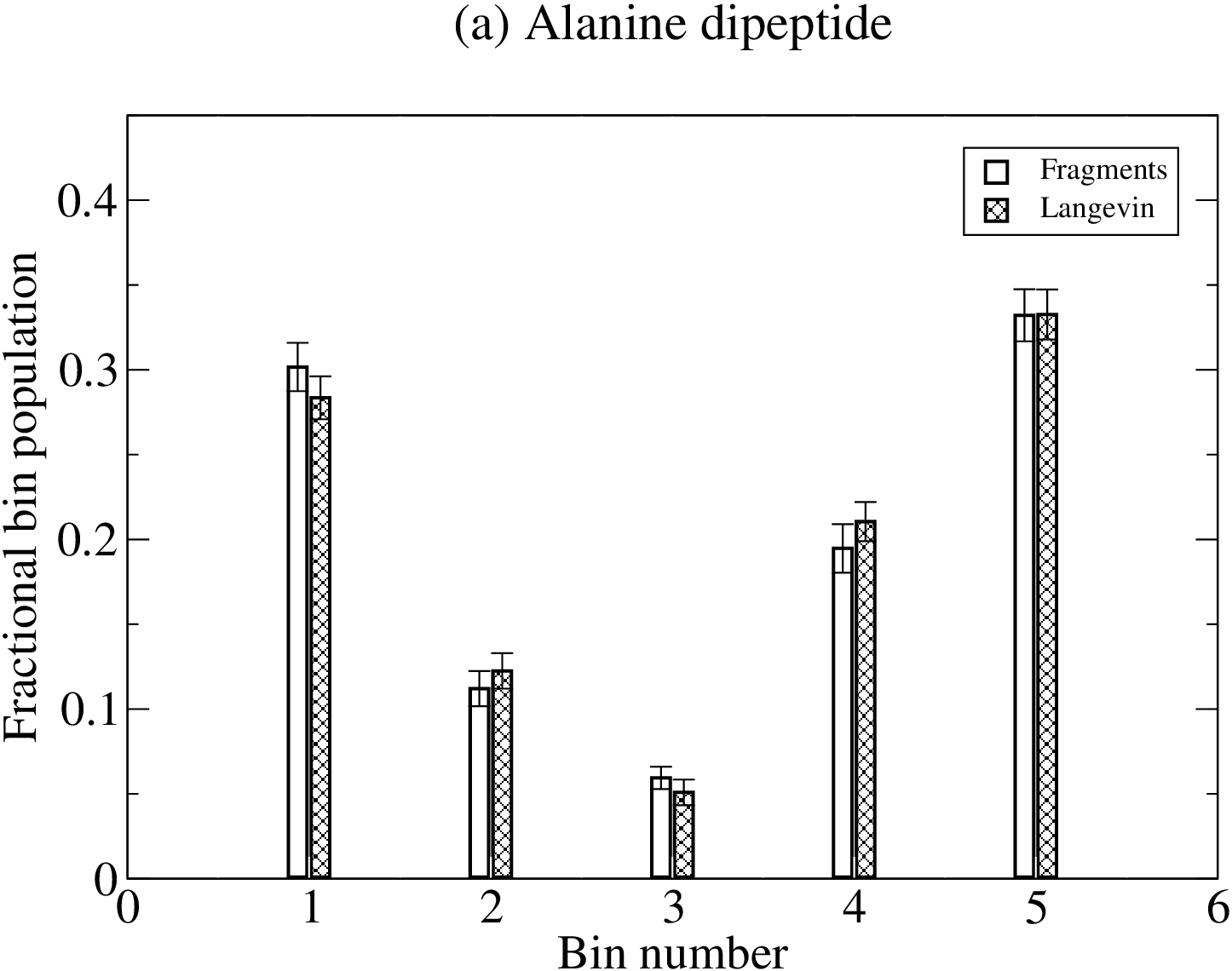}
\hfill
\includegraphics[totalheight=2.25in]{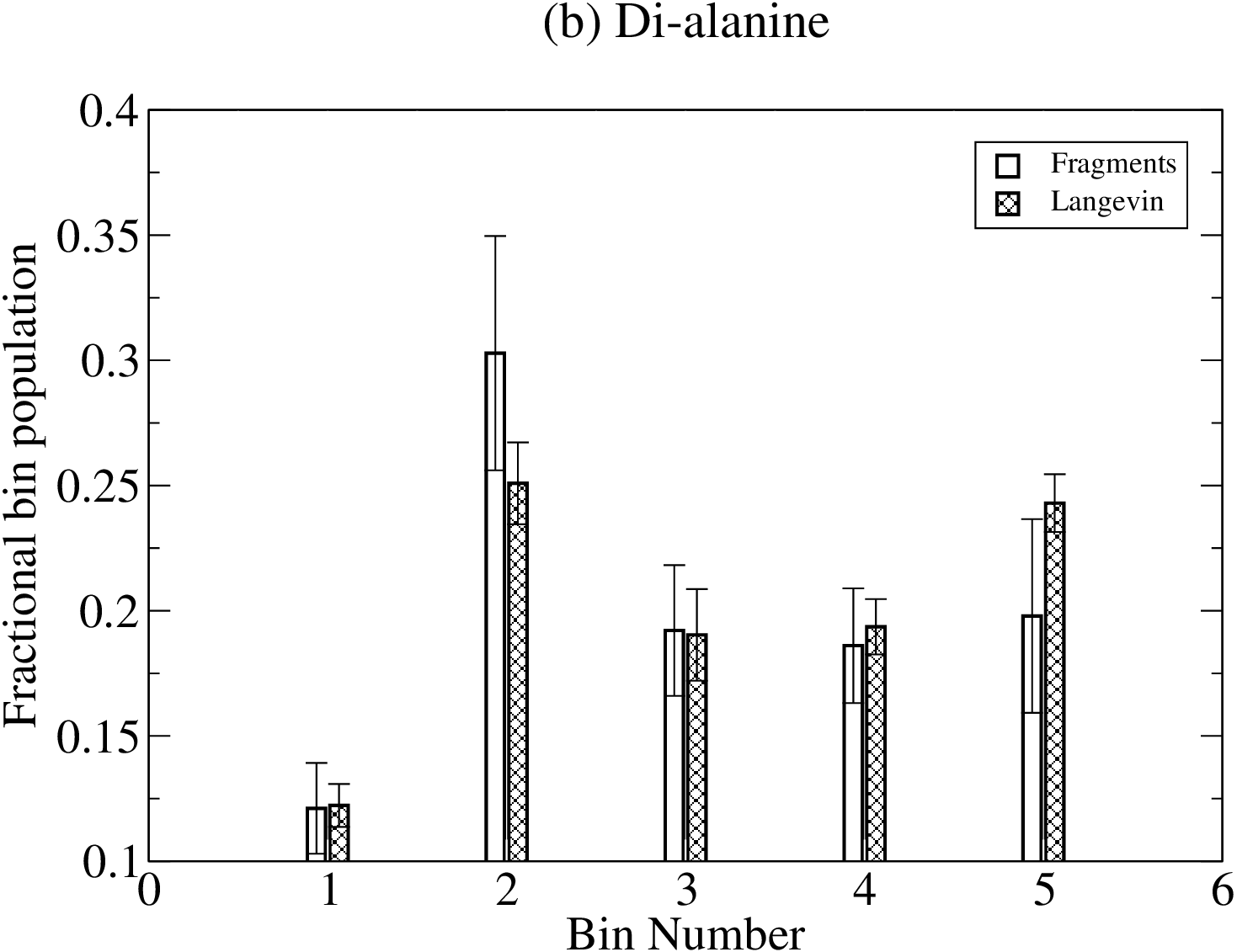}
\\
$\;$ \\
\includegraphics[totalheight=2.25in]{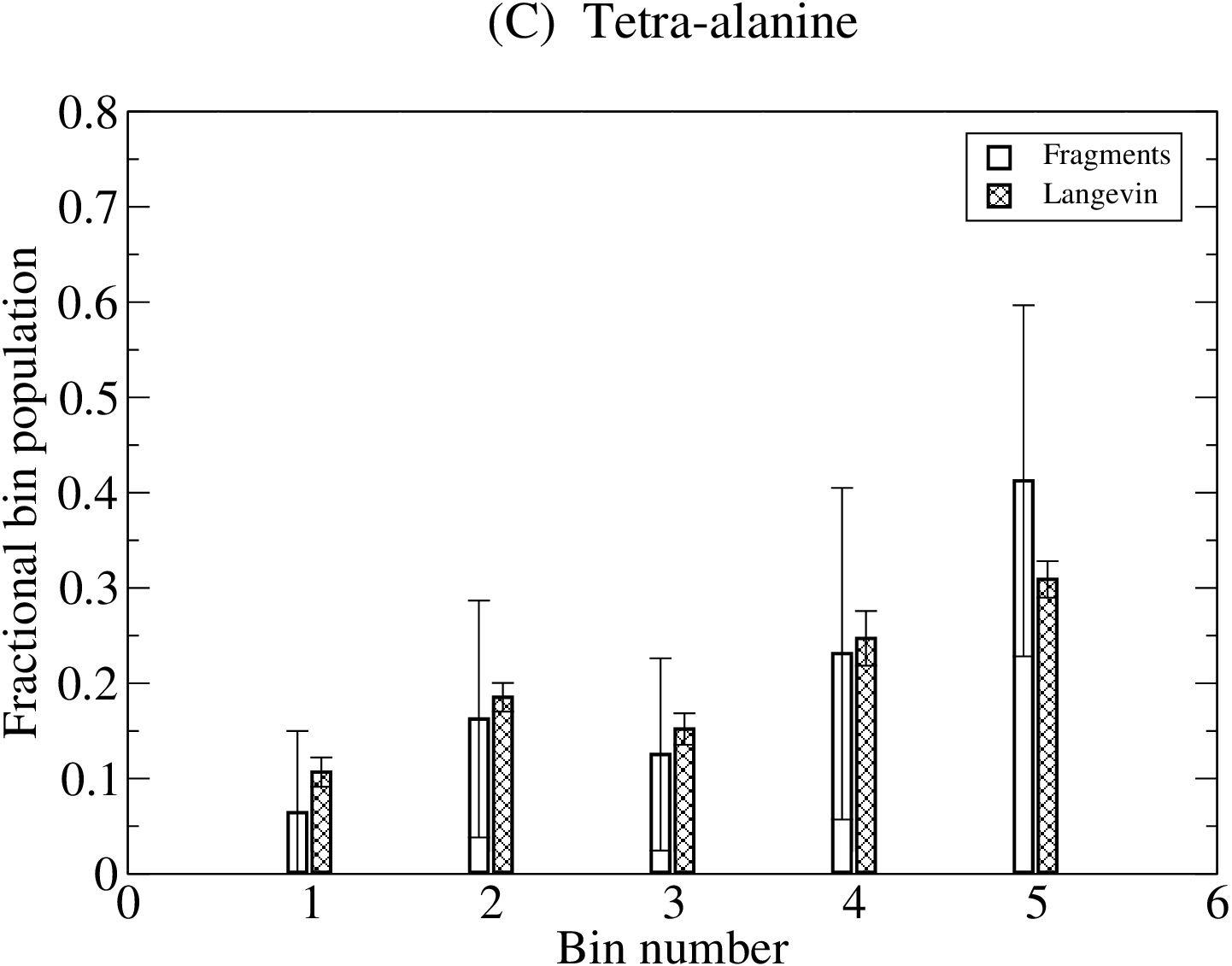}
\caption{
\label{fig:equil}
}
\end{figure}

Our free energy computation produces an equilibrium ensemble through repeated resampling procedures at each stage, as explained in the Methods section (Sec.\ \ref{sec:equil}).
As a further check on our data, we compare the equilibrium ensembles generated from our fragment combination procedure to those produced by long Langevin simulations performed in Tinker.~\cite{Ponder-1987}
The results, shown in Fig.\ \ref{fig:equil}(a), indicate that our computation is indeed producing correct equilibrium ensembles.
The graph shows the populations of different regions of configuration space, which was divided up using a Voronoi procedure explained above (Sec.\ \ref{sec:method-check}).
The alanine dipeptide equilibrium distribution was generated from the three-fragment protocol, and the 1 $\mu$sec. Langevin simulation (20*50 nsec) was performed in Tinker using a friction constant of 10.0 $ps^{-1}$ at $T=298K$.

\vspace *{1cm}

\subsection{Di-alanine}
Using the same libraries as for the alanine monomer above, we now calculate the absolute configurational free energy for the di-alanine peptide (Ace-(Ala)$_2$-Nme).
The staging used is described in Eq.\ (\ref{four-stages}), which corresponds to the following free energy terms for use in Eq.\ (\ref{f-staged}):
\begin{equation}
\begin{gathered}
\fref \equiv 0 \\
\df_1 = \face + 2 \cdot \fala + \fnme \\
\df_2 = \dfca \\
\df_3 = \dfaa \\
\df_4 = \dfan \\
\df_{4 \arr \mathrm{phys}} = \dfnon \; .
\end{gathered}
\label{da-four}
\end{equation}
It is important to note that $\dfaa$ for di-alanine differs in principle from the term with the same name in Eq.\ (\ref{ad-three}) for alanine-dipeptide because the prior ensemble is different.
This point was explained in Sec.\ \ref{sec:neighbors}.
However, in practice, the difference among the three systems is not statistically significant.

The free energy values are once again calculated with  high precision: fluctuations are a fraction of one kcal/mole.
The data for all free energy terms is given Table \ref{tab:di-tetra}, where we see a significant change in the $\dfnon$ term, reflecting the increased number of attractive interactions in this larger molecule (compared to alanine dipeptide).

Similarly, the agreement among bin populations for di-alanine in Fig.\ \ref{fig:equil}(b) is excellent, which provides an independent reason for having confidence in the free energy results.
The Langevin simulation for di-alanine was performed with exactly the same parameters as for alanine-dipeptide.

\subsection{Tetra-alanine}
Our results are of high precision ($\sim0.1$ kcal/mole standard deviation) even for tetra-alanine.
The staging follows our standard procedure, with the only subtlety in the present case is that the addition of every Ala residue is different, because the ``growing'' ensemble is different in every case.
Thus we consider the first alanine (Ala1) separate from the second (Ala2), and so on.

\begin{equation}
\begin{gathered}
\fref \equiv 0 \\
\df_1 = \face + 4 \cdot \fala + \fnme \\
\df_2 = \dfcao \\
\df_3 = \dfaat{1}{2} \\
\df_4 = \dfaat{2}{3} \\
\df_5 = \dfaat{3}{4} \\
\df_6 = \dfafn \\
\df_{6 \arr \mathrm{phys}} = \dfnon \; .
\end{gathered}
\label{ta-six}
\end{equation}

Our data for each of these terms is given in Table \ref{tab:di-tetra}.
Although the different alanine additions are based on different ensembles, the results show they are statistically indistinguishable in this case.
However, the ``non-bonded'' term $\dfnon$ again is significantly different from the previous molecules, as expected.

In comparing the equilibrium distributions from fragment combination and Langevin simulation, once again there is good statistical agreement.
For Langevin simulation of tetra-alanine, all parameters were set as before, except for a friction constant of 5.0 $ps^{-1}$, which does not alter the equilibrium distribution. 
The contrast between the large fluctuations in the bin populations $p_i$ and the high precision of  $\fphys$ in Table II  reveals an important lesson: sampling is harder than free energy calculation.
It is also noteworthy that we have already achieved significant efficiency improvements for fragment-based equilibrium sampling, beyond what is reported here, which will be reported in future work.~\cite{growth-2009}

\subsection{Timing and memory usage}
The calculations were reasonably inexpensive, taking 20 minutes for alanine dipeptide, 30 minutes for di-alanine, and 50 minutes
for tetra-alanine using one processor of an Intel Xeon 3.20 GHz machine. Concerning memory, a single library containing 10,000 configurations requires 11 MB for Ala, 12 MB for Ace-Ala complex and 5.7 MB for Ace and Nme.
An interaction table containing $10^8$ pair-wise
interactions uses 1.3 GB.

\section{Discussion}

\subsection{The overall strategy and results}
Overall, the precision of our free energy estimates was very high, which we
attribute to two related factors.  First, our ensembles in the reference and
intermediate stages were of good statistical quality --- i.e., characterized by a
large effective sample size (data not shown).~\cite{Lyman-2007}  Second, there was good
overlap between the stages, which indeed contributed to maintaining the sample size
throughout the stages.  The overlap is present by design, as interactions were
always \emph{added} between stages.  The addition of interactions or, equivalently,
correlations among degrees of freedom is guaranteed to reduce the entropy.~\cite{cover-book} This progressive narrowing of
configuration space is consistent with Kofke's proposal to calculate free energy
differences in the ``insertion'' direction.~\cite{Peter1997,Kofke1998} For larger systems, however, one
expects limitations to maintaining the sample size using the present protocol, as
explained below.

\subsection{Application of fragment combination for estimating relative protein-ligand affinities}
Because the fragment combination procedure can be applied to fragments of small molecules, and not just to peptides as in the present report, the approach can be applied to calculate approximate relative affinities.  That is, one can grow a ligand into the binding pocket of a protein receptor and calculate its free energy.  A number of different approximations can be imagined.  Most simply, the receptor can be held rigid and the ligand grown in the fields (van der Waals and electrostatic) of the receptor.  In a better approximation, the binding-site side-chains can be grown along with the ligand.  One can expect affinities based on such free energy calculations to be superior to their docking counterparts because entropy is included.  To produce a relative affinity estimate between two ligands, the respective solvation terms would need to be included as usual.~\cite{Chen-2004}

\subsection{Efficiency of fragment combination for equilibrium sampling}
As we have already noted, the fragment combination protocol we have described produces equilibrium ensembles simultaneously with free energy estimates.  
It is natural to wonder whether such ensembles are produced more efficiently than by standard dynamics simulation  especially given that small peptides have been shown to have multi-nanosecond relaxation times.~\cite{Lyman-2007}  
In fact, as we will carefully document in a separate publication,~\cite{growth-2009} fragment combination can lead to sampling that is faster by several orders of magnitude.
However, somewhat more sophisticated resampling schemes and different fragment sizes are useful in reaching the highest levels of efficiency, as will be reported.

\subsection{Use of implicit solvent models}
It is interesting and important to consider the additional costs which would be entailed by using a standard implicit solvent model, such as GBSA.~\cite{Qiu-1997}  First, both the libraries and the interaction tables would need to be regenerated using the implicit solvent model.  Although this could take several weeks of single-CPU time, it needs to be done only a single time for a given model.  The second cost is for additional solvent calculations not included in the libraries and tables. We hope to report on the staging and computational expense in a forthcoming publication.

\subsection{Relaxation simulations for large systems}
In the protocol employed for this study, the equilibrium ensemble generated at one stage, say $j$, is used to calculate the incremental free energy difference to the next stage, $j+1$.  To continue the process, the ensemble at stage $j+1$ is produced by resampling ensemble $j$ as described in the Methods section.  However, it is possible the resampled ensemble will contain a small number of distinct configurations in an important part of configuration space.  Such configurations will have high weight prior to resampling and thus the problem can be diagnosed by noting whether any configurations are resampled multiple times.  Clearly, duplicated configurations will not be statistically independent and lead to increased statistical error in free energy estimates.

One solution to this problem would be to relax duplicated configurations --- i.e. to perform short equilibrium simulations to create distinct configurations. The statistical justification of such an approach is somewhat technical~\cite{Neal-2001} and will be described in future work as required.

\subsection{Alternative staging using partial interactions}
Additional incremental stages can be added by considering only subsets of interactions.  For instance, in the case of di-alanine (Ace-(Ala)$_2$-Nme) which is composed of four fragments (A, B, C, D), there are three sets of non-bonded interactions: AC, AD, and BD.  Our present implementation adds all three in a single stage, but they could be added one at a time.  Undoubtedly, in larger systems, such finer staging will be necessary and probably should be required by relaxation of duplicated configurations as just described.

\section{Summary and Conclusions}
We have extended our earlier work on computation of absolute free energies for molecular systems ~\cite{Zuckerman-2006d} in two ways: (i) by staging the calculation using molecular fragments; and (ii) by pre-calculating ``statistical libraries'' of fragment configurations, intra-fragment energies, and inter-fragment interactions.
For a series of test systems --- the alanine monomer, dimer, and tetramer --- we were able to compute extremely precise free energies, with fluctuations $\ll$ 1 kcal/mole.
The calculations were quite fast, furthermore, with the slowest requiring less than an hour of single-processor computer time.
The speed results from employing (infinitely re-usable) pre-calculated library configurations and interactions, which pre-empt expensive energy calculations.
Our statistical libraries of amino-acid and capping-group fragments are available on our website (www.ccbb.pitt.edu/Zuckerman).

A future application of potential importance in computational biochemistry is the estimation of binding affinities of small molecules to proteins.
We hope to develop fragment libraries suitable for small molecules --- following ideas developed long ago~\cite{Leach-book,Karplus-1991} --- to be used in computing free energies within a potentially flexible protein binding site.
Protein flexibility could be included using the side-chains of the binding site as fragments in the calculations.

Another application closely related to the present report is the use of library-based fragment combination for equilibrium sampling.
While the calculations reported here already yield equilibrium ensembles, we are actively studying more efficient schemes based on advanced resampling approaches~\cite{Liu-book, Fearnhead-2003} and a range of fragment sizes.
This work will be reported in a separate publication.~\cite{growth-2009}

\section*{Acknowledgments}
We greatly appreciate the advice of Divesh Bhatt, Ying Ding, Marty Ytreberg, and Bin Zhang, as well as the financial support of the NIH (Grants GM076569 and GM070987) and NSF (Grant MCB-0643456).
%\bibliographystyle{h-physrev3}
%\bibliography{free-energy-paper} 

\begin{thebibliography}{10}

\bibitem{Frenkel-book}
D.~Frenkel and B.~Smit,
\newblock {\em Understanding Molecular Simulation} (Academic Press, San Diego,
  1996).

\bibitem{Stoessel-1990}
J.~P. Stoessel and P.~Nowak,
\newblock Macromol. {\bf 23}, 1961 (1990).

\bibitem{Zuckerman-2006d}
F.~M. Ytreberg and D.~M. Zuckerman,
\newblock J. Chem. Phys. {\bf 124}, 104105 (2006).

\bibitem{Huang-2006}
L.~Huang and D.~E. Makarov,
\newblock J. Chem. Phys. {\bf 124}, 64108 (2006).

\bibitem{Meirovitch-1992}
H.~Meirovitch,
\newblock J. Chem. Phys. {\bf 97}, 5803 (1992).

\bibitem{Meirovitch-1999}
H.~Meirovitch,
\newblock J. Chem. Phys. {\bf 111}, 7215 (1999).

\bibitem{Cheluvaraja-2004}
S.~Cheluvaraja and H.~Meirovitch,
\newblock Proc. Natl. Acad. Sci. U. S. A. {\bf 101}, 9241 (2004).

\bibitem{White-2004a}
R.~P. White and H.~Meirovitch,
\newblock Proc. Natl. Acad. Sci. U. S. A. {\bf 101}, 9235 (2004).

\bibitem{Gilson-1997}
M.~S. Head, J.~A. Given, and M.~K. Gilson,
\newblock J. Phys. Chem. B {\bf 101}, 1609 (1997).

\bibitem{Chang-2003}
C.-E. Chang, M.~J. Potter, and M.~K. Gilson,
\newblock J. Phys. Chem. B {\bf 107}, 1048 (2003).

\bibitem{Chang-2004}
C.~Chang and M.~Gilson,
\newblock J. Am. Chem. Soc. {\bf 126}, 13156 (2004).

\bibitem{Chen-2004}
C.~C. Chen, W. and M.~Gilson,
\newblock Biophys. J {\bf 87}, 3035 (2004).

\bibitem{Go-1969}
N.~Go and H.~Scheraga,
\newblock J. Chem. Phys. {\bf 51}, 4751 (1969).

\bibitem{Rosenbluth-1955}
M.~N. Rosenbluth and A.~W. Rosenbluth,
\newblock J. Chem. Phys. {\bf 23}, 356 (1955).

\bibitem{Wall-1959}
F.~T. Wall and J.~J. Erpenbeck,
\newblock J. Chem. Phys. {\bf 30}, 634 (1959).

\bibitem{Meirovich-1982}
H.~Meirovich,
\newblock J. Phys. A: Math. Gen. {\bf 15}, L735 (1982).

\bibitem{Meirovich-1985}
H.~Meirovich,
\newblock Phys. Rev. A {\bf 32}, 3699 (1985).

\bibitem{Garel-1990}
T.~Garel and H.~Orland,
\newblock J. Phys. A: Math. Gen. {\bf 23}, L621 (1990).

\bibitem{Garel-1991}
T.~Garel, J.~C. Niel, H.~Orland, J.~Smith, and B.~Velikson,
\newblock J. Chem. Phys {\bf 88}, 2479 (1991).

\bibitem{Velikson-1992}
B.~Velikson, T.~Garel, J.~C. Niel, H.~Orland, and J.~C. Smith,
\newblock J. Comput. Chem. {\bf 13}, 1216 (1992).

\bibitem{Bascle-1993}
J.~Bascle, T.~Garel, H.~Orland, and B.~Velikson,
\newblock Biopolymers {\bf 33}, 1843 (1993).

\bibitem{Grassberger-1993a}
P.~Grassberger,
\newblock J. Phys. A: Math. Gen. {\bf 26}, 2769 (1993).

\bibitem{Grassberger-1995}
P.~Grassberger and R.~Hegger,
\newblock J. Phys.- Condens. Mat. {\bf 7}, 3089 (1995).

\bibitem{Grassberger-1997}
P.~Grassberger,
\newblock Phys. Rev. E {\bf 56}, 3682 (1997).

\bibitem{Bastolla-1998}
U.~Bastolla, H.~Frauenkron, E.~Gerstner, P.~Grassberger, and W.~Nadler,
\newblock Proteins {\bf 32}, 52 (1998).

\bibitem{Liu-2002}
J.~L. Zhang and J.~S. Liu,
\newblock J. Chem. Phys. {\bf 117}, 3492 (2002).

\bibitem{Liu-2007}
J.~Zhang, M.~Lin, R.~Chen, J.~Liang, and J.~S. Liu,
\newblock Proteins {\bf 66}, 61 (2007).

\bibitem{Gibson-1987}
K.~D. Gibson and H.~A. Scheraga,
\newblock J. Comput. Chem. {\bf 8}, 826 (1987).

\bibitem{Karplus-1991}
A.~Miranker and M.~Karplus,
\newblock Proteins {\bf 11}, 29 (1991).

\bibitem{Leach-book}
A.~R. Leach,
\newblock {\em Molecular Modelling: Principles and Applications} (Pearson
  Education Limited, 2001).

\bibitem{lbmc-2008}
A.~B. Mamonov, D.~Bhatt, D.~J. Cashman, and D.~M. Zuckerman,
\newblock submitted .

\bibitem{Wall-1957}
F.~T. Wall, R.~J. Rubin, and L.~M. Isaacson,
\newblock J. Chem. Phys. {\bf 27}, 186 (1957).

\bibitem{Alexandrowicz-1969}
Z.~Alexandrowicz,
\newblock J. Chem. Phys. {\bf 51}, 561 (1969).

\bibitem{Macedonia-1999}
M.~D. Macedonia and E.~J. Maginn,
\newblock Mol. Phys. {\bf 96}, 1375 (1999).

\bibitem{Baker-2004}
C.~A. Rohl, C.~E.~M. Strauss, K.~M.~S. Misura, and D.~Baker,
\newblock Method Enzymol. {\bf 383}, 66 (2004).

\bibitem{Jorgensen-1996}
W.~L. Jorgensen, D.~S. Maxwell, and J.~Tirado-Rives,
\newblock J. Am. Chem. Soc. {\bf 118}, 11225 (1996).

\bibitem{Liu-book}
J.~S. Liu,
\newblock {\em Monte Carlo strategies in scientific computing} (Springer,
  2004).

\bibitem{Zwanzig-1954}
R.~W. Zwanzig,
\newblock J. Chem. Phys. {\bf 22}, 1420 (1954).

\bibitem{Peter1997}
D.~A.~K. Peter and T.~Cummings,
\newblock Mol. Phys. {\bf 92}, 973 (1997).

\bibitem{Kofke1998}
D.~A. Kofke and P.~T. Cummings,
\newblock Fluid Phase Equilibr. {\bf 150-151}, 41  (1998).

\bibitem{Ponder-1987}
J.~W. Ponder and F.~M. Richard,
\newblock J. Comput. Chem. {\bf 8}, 1016 (1987).

\bibitem{Zuckerman-2006b}
E.~Lyman and D.~M. Zuckerman,
\newblock Biophys. J. {\bf 91}, 164 (2006).

\bibitem{Lyman-2007}
E.~Lyman and D.~M. Zuckerman,
\newblock J. Phys. Chem. B {\bf 111}, 12876 (2007).

\bibitem{Voronoi-1907}
G.~Voronoi,
\newblock Journal für die Reine und Angewandte Mathematik. {\bf 133}, 97
  (1907).

\bibitem{growth-2009}
X.~Q. Zhang, A.~B. Mamonov, and D.~M. Zuckerman,
\newblock in progress  (2009).

\bibitem{cover-book}
T.~M. Cover and J.~A. Thomas,
\newblock {\em Elements of Information Theory (2nd Edition)} (Wiley, 2006).

\bibitem{Qiu-1997}
D.~Qiu, P.~S. Shenkin, F.~P. Hollinger, and W.~C. Still,
\newblock J Phys. Chem. A {\bf 101}, 3005 (1997).

\bibitem{Neal-2001}
R.~M. Neal,
\newblock Stat. comput. {\bf 11}, 125 (2001).

\bibitem{Fearnhead-2003}
P.~Fearnhead and P.~Clifford,
\newblock J. R. Stat. Soc. B {\bf 65}, 887 (2003).

\end{thebibliography}

\newpage

Figure captions:

Figure 1: Stages for calculating the absolute free energy of a molecule by combining three fragments, based on Eq.\ (\ref{three-stages}).
Connecting lines schematize full interactions between fragments, including both bonded and non-bonded atomistic terms.
(a) The first intermediate stage comprises non-interacting fragments, but includes all interactions \emph{internal} to each fragment. 
(b) The second stage adds interactions among the atoms of fragments A and B, while (c) the third stage does the same for fragments B and C. 
(d) In the final stage, representing the desired free energy $\fphys$, all interactions are added, including among non-sequential fragments and possibly including an implicit solvent model.

Figure 2: Stages used in the free energy calculation of a four-fragment molecule, corresponding to Eq.\ (\ref{four-stages}).
The initial stages proceed in analogy to Fig.\ \ref{fig:abc}, with pair-wise interactions added one at a time for neighboring (``bonded'') fragments.
In the final stage, \emph{all} remaining interactions are added.
Other, more incremental staging schemes are possible, but were not necessary in the present study.

Figure 3: Comparison of equilibrium distributions from fragment combination and Langevin simulation.The graphs show the fractional population in different regions of configuration space, as described in Sec.II K.
Three peptides are considered: (a) alanine dipeptide, (b) di-alanine, and (c) tetra-alanine.
The error bars for both the fragment combination and Langevin results reflect twice the standard deviations among 20 independent simulations, roughly a 95\% confidence interval. Each Langevin simulation was 50 nsec long.
The statistical agreement is good in every case.\\
\newpage
\begin{table}
\begin{tabular}{rc|rc}
\hline \hline
\multicolumn{4}{c}
{Alanine dipeptide free energy terms from Eqs.\ (\ref{ad-three}) and (\ref{ad-two})} 
\\
\hline
\multicolumn{2}{c|}{Three Fragments} &
\multicolumn{2}{c}{Two Fragments}
\\
\hline
Term & $\;\;\;\;$ Estimate [kcal/mol] $\;\;\;\;$ 
& Term & Estimate [kcal/mol]
\\
\hline
$\face$  &  14.783(0.003)  &  $\faceala$  &  47.311(0.027)
\\
$\fala$  &  33.326(0.015)  &  $\fnme$     &  16.574(0.003)
\\
$\fnme$  &  16.574(0.003)  &  $\;\;\;\;\dfcan$    &  $-$0.792(0.002)
\\
$\dfca$  &  $-$0.801(0.002)  &&
\\
$\dfan$  &  $-$0.499(0.007)  &&
\\
$\dfnon$ &  $-$0.285(0.008)  &&
\\
\hline
$\fphys$ &  63.098(0.015)  &  $\fphys$     &  63.093(0.028)
\\
\hline \hline
\end{tabular}
\caption{
\label{tab:ala-dipep}
Comparison between the absolute free energy for alanine dipeptide estimate using two different fragmentation schemes.
Our ``standard'' three-fragment decomposition (Ace, Ala, Nme) is compared to a two-fragment grouping (Ace-Ala, Nme). 
The table gives free energy values in kcal/mole, as well as two standard deviations (in parentheses) based on 20 independent calculations.
}
\end{table}

\newpage
\begin{table}
\begin{tabular}{rc|rc}
\hline \hline
\multicolumn{4}{c}
{Free energy terms for di-alanine from Eq.\ (\ref{da-four}) and for tetra-alanine from Eq.\ (\ref{ta-six})} 
\\
\hline
\multicolumn{2}{c|}{Di-alanine} &
\multicolumn{2}{c}{Tetra-alanine}
\\
\hline
Term & $\;\;\;\;$ Estimate [kcal/mol] $\;\;\;\;$ 
& Term & Estimate [kcal/mol]
\\
\hline
$\face$  &  14.783(0.003)    &  $\face$         &  14.783(0.003)
\\
$\fala$  &  33.326(0.015)    &  $\fala$         &  33.326(0.015)
\\
$\fnme$  &  16.574(0.003)    &  $\fnme$         &  16.574(0.003)
\\
$\dfca$  &  $-$0.801(0.002)  &  $\;\;\;\;\dfcao$  &  $-$0.801(0.002)
\\
$\dfaa$  &  $-$0.771(0.014)  &  $\dfaat{1}{2}$  &  $-$0.774(0.013)
\\
$\dfan$  &  $-$0.499(0.013)  &  $\dfaat{2}{3}$  &  $-$0.774(0.012)
\\
$\dfnon$ &  $-$0.809(0.031)  &  $\dfaat{3}{4}$  &  $-$0.771(0.014)
\\
&                            &  $\dfafn$        &  $-$0.498(0.009)
\\
&                            &  $\dfnon$        &  $-$1.986(0.284)
\\
\hline
$\fphys$ &  95.128(0.057)    &  $\fphys$        &  159.057(0.293)
\\
\hline \hline
\end{tabular}
\caption{
\label{tab:di-tetra}
Free energy terms used in calculating the absolute free energy for di-alanine and tetra-alanine.
The table gives free energy values in kcal/mole, as well as two standard deviations (in parentheses) based on 20 independent calculations.
}
\end{table}

\end{document}